\documentstyle[12pt,epsf]{article}

\newcommand{\bmat}{\left(\begin{array}}
\newcommand{\emat}{\end{array}\right)}

\def\yzero{\smash{\hbox{$y\kern-4pt\raise1pt\hbox{${}^\circ$}$}}}
\def\p{\partial}

\def\p{\tilde p}

\def\-{\hphantom{-}}

\def\s2{\frac{1}{\sqrt2}}

\def\beq{\begin{equation}}
\def\eeq{\end{equation}}
\def\beqa{\begin{eqnarray}}
\def\eeqa{\end{eqnarray}}

\def\IF{\relax{\rm I\kern-.18em F}}
\def\II{\relax{\rm I\kern-.18em I}}
\def\IP{\relax{\rm I\kern-.18em P}}
\def\inbar{\vrule height1.5ex width.4pt depth0pt}
\def\IC{\relax\hbox{\kern.25em$\inbar\kern-.3em{\rm C}$}}
\def\IR{\relax{\rm I\kern-.18em R}}

\def\Dsl{\,\raise.15ex\hbox{/}\mkern-13.5mu D} 

\newcommand{\drawsquare}[2]{\hbox{%
\rule{#2pt}{#1pt}\hskip-#2pt
\rule{#1pt}{#2pt}\hskip-#1pt
\rule[#1pt]{#1pt}{#2pt}}\rule[#1pt]{#2pt}{#2pt}\hskip-#2pt
\rule{#2pt}{#1pt}}

\newcommand{\fund}{\raisebox{-.5pt}{\drawsquare{6.5}{0.4}}}
\newcommand{\antifund}{\overline{\fund}}

\topmargin
-1.5cm
\textwidth
15.5cm
\textheight
24cm
\oddsidemargin
0.7cm
\evensidemargin
1.2cm

\begin{document}

\makeatletter
\@addtoreset{equation}{section}
\makeatother
\renewcommand{\theequation}{\thesection.\arabic{equation}}
\pagestyle{empty}
\rightline{IFT UAM-CSIC/03-27}
\rightline{\tt hep-th/0307196}
\vspace{0.5cm}
\begin{center}
\LARGE{From UV/IR mixing to closed strings\\[20mm]}
\large{
Esperanza L\'opez\\[5mm]}
\small{
Departamento de F\'{\i}sica Te\'orica C-XI
and Instituto de F\'{\i}sica Te\'orica  C-XVI\\[-0.3em]
Universidad Aut\'onoma de Madrid,
Cantoblanco, 28049 Madrid, Spain\\
Esperanza.Lopez@uam.es\\[4mm]}
\small{\bf Abstract} \\[7mm]
\end{center}

\begin{center}
\begin{minipage}[h]{14.0cm}

It was shown in hep-th/0301099 that the leading UV/IR
mixing effects in noncommutative gauge theories on D-branes
are able to capture information about the
closed string spectrum of the parent string theory.
The analysis was carried out for D-branes on nonsupersymmetric
$\mathbf{C}^3/\mathbf{Z}_N$ orbifolds of Type IIB.
In this paper we consider D-branes on twisted circles 
compactifications of Type II string theory. We find 
that the signs of the leading UV/IR mixing effects know about the 
mass gap between the lowest modes in NSNS and RR closed 
string towers. Moreover, the relevant piece of the field theory 
effective action can be reproduced purely
in the language of closed strings. Remarkably this approach
unifies in a single structure, that of a closed string exchange
between D-branes, both the leading planar and nonplanar
effects associated to the absence of supersymmetry.

\end{minipage}
\end{center}
\newpage
\setcounter{page}{1}
\pagestyle{plain}
\renewcommand{\thefootnote}{\arabic{footnote}}
\setcounter{footnote}{0}

\section{Introduction}

The study of noncommutative field theories has attracted much interest
in the last years because noncommutativity is expected to capture basic 
aspects of the sort distances behavior of gravity \cite{Doplicher:1994tu}. 
This point of view is supported by the central role that D-branes 
play in the matrix model description of M-theory \cite{Banks:1996vh},
since D-branes provide a fundamentally noncommutative description of 
space-time \cite{Witten:1995im}. 

In order to explore the consequences of noncommutativity, most works 
have considered the simple deformation of $\mathbf{R}^n$
\beq
[x^\mu , x^\nu]=i \theta^{\mu \nu} \, ,
\label{noncom}
\eeq
where $\theta^{\mu \nu}$ is an antisymmetric matrix with constant entries.
The reason to focus in (\ref{noncom}) is that explicit calculations 
can be done for field theories living on such spaces. (\ref{noncom}) 
implies uncertainty relations in space-time and therefore 
the nondecoupling of ultraviolet and infrared degrees of freedom.
This new behavior has drastic consequences for the dynamics of the
theory and in particular for its nonplanar sector: It was shown in 
\cite{Minwalla:1999px} that UV/IR mixing leads
to the appearance of new infrared divergences in nonplanar graphs, 
whose origin is in the integration to high momenta in loops. 

The relations (\ref{noncom}) are naturally realized on the world-volume
of D-branes in a constant B-field background, where 
$\theta^{\mu \nu} \sim 1/B_{\mu \nu}$ \cite{Douglas:1997fm}. Nonplanar 
field theory diagrams can then be related to nonplanar string diagrams.
This suggests an important role of closed strings in the understanding 
of UV/IR mixing \cite{Minwalla:1999px,Arcioni:2000bz}. 
More generically, it is natural to expect that if 
(\ref{noncom}) captures relevant aspects of quantum gravity its effects 
will know about the closed string sector (although the decoupling of
closed strings {\it does not} fail in the noncommutative field theory 
limit \cite{Gomis:2000bn,Liu:2000qh}). In this line, a remarkable 
relation between UV/IR mixing effects and {\it properties} of the closed 
string spectrum was uncovered in \cite{Armoni:2003va}. 
The leading IR divergences due to UV/IR mixing strongly modify 
the dispersion relations of the theory and in some cases render the 
perturbative vacuum unstable \cite{Minwalla:1999px}. 
A one to one correspondence
between these noncommutative instabilities and the presence of 
closed string tachyons in the parent string theory was obtained
for gauge theories on D-branes 
at nonsupersymmetric $\mathbf{C}^3/\mathbf{Z}_N$ orbifold singularities.

${\cal N}=4$ noncommutative $U(1)$ at finite temperature
presents a non-trivial but mild version
of UV/IR mixing, where no infrared divergence develops.
In spite of that, for temperatures bigger than a critical $T_c$ 
excitations of tachyonic nature appear in the system 
\cite{Landsteiner:2001ky}. A closely related 
theory arises on D3-branes wrapped
along a Scherk-Schwarz circle. It can be seen that noncommutative 
instabilities set up well below the threshold for the appearance of 
closed string tachyons, invalidating a direct relation between both
phenomena. A motivation of this paper is to determine whether the leading
UV/IR mixing effects know about properties of the closed string spectrum 
in this more general case.

The paper is organized as follows. In section 2 we summarize the 
results obtained in \cite{Armoni:2003va} for noncommutative 
$\mathbf{C}^3/\mathbf{Z}_N$ quiver theories. In section 3 we consider 
Type II string theory compactified on twisted circle backgrounds 
\cite{Russo:2001na}, which are generalizations of a Scherk-Schwarz 
compactification. The closed string spectrum contains tachyons for
sufficiently small radius of the circle. In section 3 and 4 we analyze 
Type IIB D3-branes wrapped along the twisted circle. In section 5 we 
study Type IIA D2-branes transverse to the twisted circle. The field 
theory limit on D2-branes and D3-branes corresponds to
the string regimes with and without closed string tachyons respectively.
The sign of the leading UV/IR mixing effects determines if the associated 
correction will tend to destabilize or not the perturbative vacuum.
For the gauge theories on both D2- and D3-branes we find that these 
signs are correlated with the $($mass$)^2$ difference between the 
lowest NSNS and RR modes in each winding sector. This will 
be our proposal on how to generalize the results of \cite{Armoni:2003va}.
Notice that for $\mathbf{C}^3/\mathbf{Z}_N$ orbifolds the RR zero-point 
energy is zero and thus the sign of the gap indicates the presence or not 
of closed string tachyons. However for twisted circle compactifications 
the mass of closed strings, both in NSNS and RR sectors, is shifted by a 
nonzero winding energy.

We include in section 2 a proposal for deriving, purely in the language 
of closed strings, the piece of the gauge invariant effective action 
responsible for the instabilities \cite{Armoni:2001uw}. This 
provides a very direct relation between UV/IR mixing effects and closed 
strings. A second goal of this paper is to extend this derivation to the 
gauge theories on D2- and D3-branes at twisted circle backgrounds.
We will show in sections 4 and 5 that this is indeed
possible. All the leading UV/IR mixing effects in nonplanar 
graphs, as well as the leading planar corrections, have the structure 
of a closed string exchange. The possibility to unify planar 
and nonplanar effects in a single structure shreds light on  
the puzzling different behavior of planar and nonplanar sectors.
We present our conclusions in section 6.
In an appendix we include the derivation of some useful field theory
results.

\section{Open Wilson lines versus closed strings}

\begin{figure}
\centering
\epsfxsize=3.5in
\hspace*{0in}\vspace*{0in}
\epsffile{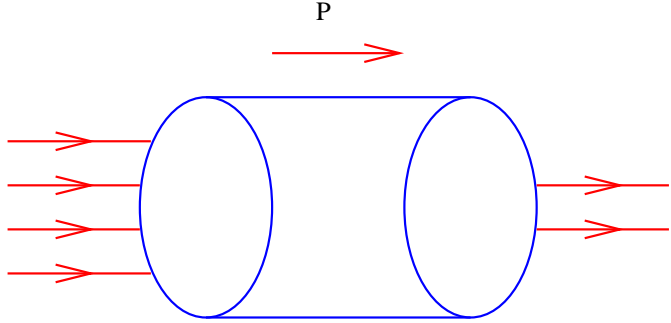}
\caption{\small Schematic representation of a non-planar graph.}
\label{channel}
\end{figure}

Non-planar graphs in noncommutative field theories posses an
intrinsic UV cutoff in the form of phases whose argument oscillates 
rapidly for high loop momenta. When the momentum exchange in 
the non-planar channel vanishes (see Fig. 1), these phases trivialize 
and the regularization stops being effective. This implies 
$\Lambda_{np} \sim {1 \over P \theta}$ and, as a consequence, the
translation of UV into IR behavior. In its most extreme form, it 
converts potential UV divergences into IR divergences 
\cite{Minwalla:1999px}. 

The leading IR divergences modify the dispersion relations of 
noncommutative field theories as follows 
\beq
E^2= {\vec p}^2 - c \; {g^2 \over \p^2} \, ,
\label{disp}
\eeq
where $g$ is the (dimensionless) coupling constant, 
$\p^\mu=\theta^{\mu \nu} p_\nu$ and $c$ is a model dependent constant. 
For non-commutative gauge theories $c\sim N_b-N_f$, with $N_b$ and $N_f$ 
the number of bosonic and fermionic degrees of freedom in the adjoint 
representation. Supersymmetric theories have a softer UV behavior and 
thus the leading IR divergences are absent: $c=0$ 
\cite{Matusis:2000jf}. When $N_b>N_f$, (\ref{disp}) turns the low 
momentum modes unstable \cite{Ruiz:2001hu,Landsteiner:2001ky}. 

In order to explore a possible connection between noncommutative instabilities 
and closed string tachyons, it proved useful in \cite{Armoni:2003va} to 
study a family of theories with a sufficiently rich phenomenology. This 
was provided by noncommutative D-branes on $\mathbf{C}^3/\mathbf{Z}_N$ 
orbifold backgrounds. 
The orbifold acts with twist $(a_1,a_2,a_3,a_4)/N$ and $(b_1,b_2,b_3)/N$ 
on $SO(6)$ spinors and vectors respectively. The integers $a_\alpha$ are 
subject to $\sum a_\alpha=0({\rm mod} N)$ and are related to
$b_l$ by $b_1=a_2+a_3$, $b_2=a_1+a_3$, $b_3=a_1+a_2$.
The gauge theory on $n$ D3-branes placed at the fixed 
point of the orbifold has gauge group $G=\otimes_{i=1}^N U(n_i)$, where 
$\sum n_i=n$. The coupling constants of all gauge group factors coincide.
The matter content is given by 
$(\fund_i,\antifund_{i+a_\alpha})$ Weyl fermions and $(\fund_i,
\antifund_{i+b_l})$ complex scalars. Both the gauge theory 
on the D3-branes and the closed string spectrum are supersymmetric 
if at least one $a_\alpha=0({\rm mod} N)$. 

Turning on a B-field background on two of the spatial directions of the 
D3-branes will render the world-volume noncommutative, {\it i.e.} 
$[x^1,x^2]=i \theta$. In the generic nonsupersymmetric case the 
noncommutative gauge theory 
presents a complicated pattern of UV/IR mixing effects. The leading 
infrared contribution to the nonplanar polarization tensor was calculated 
in \cite{Armoni:2003va}. As can be seen 
from Fig. 2, it affects only $U(1)_i \in U(n_i)$ degrees of 
freedom\footnote{In the absence of noncommutativity the theories we are 
considering have generically mixed anomalies, that will lead to the
massification of the $U(1)$'s through a Green-Schwarz
mechanism \cite{Ibanez:1998qp}. When $\theta \neq 0$ the situation is 
more subtle: the $U(1)$'s become massive and decouple only for $\p=0$ 
\cite{Armoni:2002fh}, while for $\p \neq 0$ the anomaly vanishes
\cite{Martin:2000qf}. We will always consider
the latter case.}
and is non-diagonal in group labels indices. However the linear 
combinations $B^{(k)}_\mu= {1 \over \sqrt{N}} \sum e^{2 \pi i {jk \over N}} 
{\rm Tr} A^{(j)}_\mu \,$ diagonalize it, with the result
\beq
\Pi^{\mu \nu}_k= 
\epsilon_k \; {g^2 \over \pi^2} {\p^\mu \p^\nu \over \p^4} \, .
\label{pior}
\eeq
The quantities $\epsilon_k$, which play an analogous role to $c$ in 
(\ref{disp}), have a simple expression in terms
of the orbifold twist parameters 
\beq
\epsilon_k = 2 \left( 1-\sum_{\alpha=1}^4 
\cos {2 \pi a_\alpha k \over N} +\sum_{l=1}^3 
\cos {2 \pi b_l k \over N} \right) \, .
\label{ep}
\eeq 

Remarkably, these quantities can be rewritten in terms of the 
masses of four low lying closed string modes in the NSNS $k^{th}$ twisted 
sector of the orbifold background
\beq
\epsilon_k=-16 \prod_{\alpha=1}^4 \sin {\pi \alpha' m_\alpha^2 \over 2} \, .
\label{emass}
\eeq
These masses satisfy the following properties: {\it i)}
among them is the lowest mode in the NSNS $k^{th}$ twisted sector;
{\it ii)} $-1 \leq \alpha' m_\alpha^2 < 2$; {\it iii)} only one
of the $m_\alpha^2$ can be negative (see \cite{Armoni:2003va} for details). 
This implies a direct relation between the sign 
of $\epsilon_k$ and the presence or not of tachyons in the $k^{th}$ twisted 
sector of the parent string theory: $\epsilon_k>0$ if the $k^{th}$ twisted
sector contains tachyons; $\epsilon_k=0$ only if the $k^{th}$ twist
is supersymmetry-preserving; 
$\epsilon_k<0$ when the twisted sector
is nonsupersymmetric but does not include tachyons. 
This last situation does not
arise in $\mathbf{C}^n/\mathbf{Z}_N$ orbifolds with $n<3$, and was the reason 
to consider orbifolds of $\mathbf{C}^3$.

\begin{figure}
\centering
\epsfxsize=2.5in
\hspace*{0in}\vspace*{0in}
\epsffile{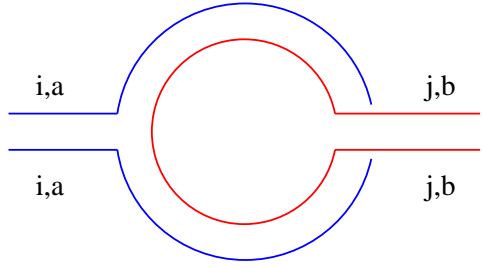}
\caption{\small Non-planar contribution to the polarization tensor.
$i,j$ are group label indices and $a,b$ gauge group indices.}
\label{nonplannar}
\end{figure}

Let us come back now to the origin of UV/IR mixing, which is in the
uncertainty relations derived from (\ref{noncom}): 
$\Delta x^1 \Delta x^2 \geq \theta$ in our case. This phenomenon is 
present in any noncommutative theory, independently whether 
there are UV or IR divergences. It would seem then that the natural degrees 
of freedom of noncommutative theories are those that made this behavior 
evident. Contrary to this expectation, noncommutative field theories 
are formulated using local fields. 
This puzzle is solved by studying the effective action of the theory.
It has been shown both for scalar \cite{Kiem:2001pw} and  
gauge theories \cite{VanRaamsdonk:2001jd,Armoni:2001uw,Kiem:2001dm}
that the 1-loop nonplanar effective action, including contributions from all 
the N-point functions, can be rewritten in terms of straight open Wilson line 
operators \cite{Ishibashi:2000hs,Rey:2000hh,Das:2000md,Gross:2000ba}. 
This result was extended 
in \cite{Kiem:2001du} to the 2-loop effective action of scalar theories. 
Straight open Wilson line operators exhibit the desired behavior
since their momentum, $p$, is correlated with their transversal extent
$\p$,
\beq
{\widetilde W}(p)\, = \, {\rm Tr} \int d^4 x \; P_\ast 
\left(e^{i\, g \int_0^1 d \sigma 
\, {\tilde p}^\mu A_{\mu}(x+{\tilde p}\, \sigma)}\right) 
\ast e^{i p x} \, .
\label{w}
\eeq
For the orbifold gauge theories above, the gauge invariant piece
of the 1-loop effective action containing (\ref{pior}) has the
simple expression \cite{Armoni:2003va}
\beq
\Delta S = {1 \over 2 \pi^2} \sum_{k=0}^{N-1} \epsilon_k
\int {d^4 p \over (2 \pi)^4} \; {1 \over {\tilde p}^4} \; 
W^{(N\!-k)}(p) \, W^{(k)}(-p) \, ,
\label{invaction}
\eeq
where $W^{(k)} = {1 \over \sqrt{N}}\sum_{j=1}^N e^{2 \pi i \, {jk \over N}} 
{\widetilde W}^{(j)}$, and ${\widetilde W}^{(j)}$ denotes (\ref{w}) with
the vector field belonging to the $j^{th}$ gauge group factor.

Several facts suggest the interpretation of (\ref{invaction}) in terms
of a closed string exchange between D-branes. (\ref{invaction}) seems 
to know about the different closed string twisted sectors since the 
quantities $\epsilon_k$ measure an independent property of each sector. 
In addition, closed string modes in the $k^{th}$ twisted sector couple to 
linear combinations of field theory operators such as those that define 
$B_\mu^k$ and $W^k$ \cite{Douglas:1996sw}. Moreover, it has been 
shown in \cite{Das:2000ur,Okawa:2000sh,Liu:2001ps} that closed strings 
couple to straight Wilson line operators on noncommutative D-branes.
We should notice that closed strings can only couple to gauge invariant 
operators and the open Wilson lines provide the basic set of such operators in 
noncommutative gauge theories \cite{Gross:2000ba}. However, the condition 
of gauge invariance only determines the separation between the start and end 
point of the Wilson line but not the path.

Reviewing \cite{Armoni:2001uw,Armoni:2003va}, we will propose now a direct 
derivation of (\ref{invaction}) in terms of a closed strings. Our motivation 
is 
twofold. We have mentioned two aspects of UV/IR mixing, the first associated 
to the physics of the non-planar sector and the appearance of an extreme
infrared behavior, the second associated to the question of what are
the natural noncommutative excitations and the role of open Wilson line 
operators. Closed strings provide a connection between these two
aspects since they couple naturally to open Wilson lines and, as we will
see, the IR divergent term $1/\p^4$ in (\ref{invaction}) can be
related to a closed string propagator.
Our second goal is to obtain an alternative understanding of the 
$\epsilon_k$'s and its relation to closed string tachyons.

In the absence of B-field, scalar closed string modes couple to the
brane tension at leading order in $\alpha'$: $T \sim 
{\rm Tr} \, {\mathbf 1}/\alpha'^2$. When $B\neq 0$ the trivial field 
theory operator ${\rm Tr} \,{\mathbf 1}$ gets promoted
to the open Wilson line operator (\ref{w}). Hence, at leading order in 
$\alpha'$, the contribution to the D3-brane effective action from the 
emission, propagation and posterior absorption of a scalar closed string 
mode $\varphi$ in the $k^{th}$ twisted sector is  
\beq
\Delta S= |D_\varphi|^2 \int {d^4 p \over (2 \pi)^4} \, 
W^{(N-k)}(p) W^{(k)}(-p) \, f({\tilde p},u) \, ,
\label{staction}
\eeq
with $D_\varphi$ a numerical factor proportional to the disk amplitude 
of $\varphi$ with no open string insertion. The function $f({\tilde p},u)$
denotes the closed string propagator
\beq
f({\tilde p},u)= {\alpha'}^{-{d+2 \over 2}} \int {d^d v \over
(2 \pi)^d} {e^{i v u} \over v^2+{\tilde p}^2+(2 \pi \alpha' m_\varphi)^2} \, .
\label{f}
\eeq
$d$ is the number of dimensions
transverse to the D-brane where the twisted field $\varphi$ can propagate: 
$d=0,2,4,6$ depending on the particular $\mathbf{C}^3/\mathbf{Z}_N$ 
orbifold. We have defined $v=2 \pi \alpha' p_\perp$, with $p_\perp$ the 
transversal momentum to the D-brane. $u$ has been introduced in order to have 
a well defined
closed string propagator, it has the interpretation of an infrared regulator 
from the point of view of the field theory. In the denominator we have used 
the relation between open ($\eta$) and closed ($g$) string metrics 
$g^{-1}=\eta^{-1}\!-\! \theta \eta \theta/(2 \pi \alpha')^2$ 
\cite{Seiberg:1999vs}, and discarded
terms suppressed by two $\alpha'$ powers; $m_\varphi$ is the mass of $\varphi$.
The factor of $\alpha'$ in front of the integral can be
obtained just by dimensional analysis. 

From (\ref{staction}) we want to extract
a contribution to the noncommutative field theory effective action.
We need to take the limit $\alpha' \rightarrow 0$. In this limit $f$ 
diverges due to the negative $\alpha'$ power in front of the integral. 
However we should notice the following.
If we had done the $\alpha' \rightarrow 0$ limit of the standard annulus 
diagram associated to each nonplanar N-point function, 
we would had obtained a result ${\cal O}({\alpha'}^0)$ whose leading
IR contribution should reproduce the corresponding term in (\ref{invaction}). 
This has been checked for the 2-point function in 
\cite{Bilal:2000bk,Gomis:2000bn}. The question is then whether
we can directly define an ${\cal O}({\alpha'}^0)$ contribution
from (\ref{staction}), regarding as artifacts other $\alpha'$ powers.
We observe that $\p$ acts as an infrared regulator for the integral in 
(\ref{f}). Therefore, for $\p \neq 0$, we can expand the integral in powers 
of $(2 \pi \alpha' m_\varphi)^2 \sim \alpha'$ to the desired order.
At ${\cal O}({\alpha'}^0)$ we obtain
\beq
f({\tilde p},u)|_{{\cal O}(\alpha'^0)}\, \sim \, \int {d^d v \over
(2 \pi)^d} {e^{i v u} \over (v^2+{\tilde p}^2)^{{d \over 2}+2}} 
\rightarrow_{u \rightarrow 0} {c \over \p^4} \, ,
\label{ffin}
\eeq
where $c$ is a constant that depends on the dimension $d$ and whose
value we will not need to determine. After removing the field theory 
infrared regulator $u$ we recover $1/\p^4$, independently of $d$. 
The possibility to relate 
$1/\p^4$ to a closed string propagator does not mean that the 
decoupling of closed strings fails in the noncommutative field theory 
limit, since the IR singularities do not have kinetic part. Hence 
they do not force the introduction of additional degrees of freedom.

In (\ref{staction}) we have considered the exchange of a single
closed string mode. Any closed string mode able to couple to $W^k$
will contribute the same $1/\p^4$ up to a numerical factor
depending on $D_\varphi$ and $m_\varphi$. The coefficients $\epsilon_k$
that appear in (\ref{invaction}) will be thus a collective effect of the 
closed string towers. 
However, we know from (\ref{emass}) that $\epsilon_k$ is determined
by a finite set of low lying string modes and, moreover, its sign 
just depends on the presence of tachyons. To reconcile these
two facts, notice that modes in the NSNS and RR sectors contribute 
with opposite sign to the exchange between D-branes. 
In addition, when the theory is supersymmetric we have 
$\epsilon_k=0$, implying that the contribution from both
towers must cancel. This suggests to consider $\epsilon_k$ as 
a measurement of the misalignment between the NSNS and RR towers. 
It seems then consistent that the presence of tachyons, 
which can only belong to the NSNS sector, is linked to the sign of the 
mentioned misalignment given that the lightest RR modes are massless. 

We would like to end this section with a comment on the cases  
with scalars in the adjoint representation (this 
situation only arises if some $b_l=0$). 
It was shown in \cite{Das:2000ur,Okawa:2000sh,Liu:2001ps} that in those 
cases closed string modes couple to generalized Wilson line operators 
\beq
{\widetilde W}(p,v)\, = \, {\rm Tr} \int d^4 x \; P_\ast 
\left(e^{i\, g \int_0^1 d \sigma \big(
\, {\tilde p}^\mu A_{\mu}(x+{\tilde p}\, \sigma) + 
v_a \phi_a(x+ {\tilde p} \sigma) \big)}\right) 
\ast e^{i p x} \, ,
\label{Ws}
\eeq
where $v_a$ label as before the momentum in the directions transverse to 
the D3-branes and $\phi_a$ denote the adjoint (real) scalars. In this 
case the closed string derivation of the effective action we have 
proposed suggests to extent (\ref{invaction}) to
\beq
\Delta S = {1 \over 2 \pi^2 c} \sum_{k=0}^{N-1} \epsilon_k
\int {d^4 p \over (2 \pi)^4} {d^d v \over (2 \pi)^d }
\; W^{(N\!-k)}(p,v) \, W^{(k)}(-p,-v) {e^{ivu} \over 
(v^2+\p^2)^{{d \over 2}+2}} \, , 
\label{actsc}
\eeq
with $c$ the same constant as in (\ref{ffin}). Adjoint scalars, as gauge 
bosons, get pole-like IR corrections to their self-energy. They can be 
obtained from (\ref{actsc}) by expanding $W^k$ to linear order in 
${\rm Tr} \phi_a$ using\footnote{The integral (\ref{ffin}) for $d=2,4,6$
can be performed explicitly resulting in all cases proportional to 
${u^2 \over \p^2} K_2(u \p)$,
with $K_2$ a modified Bessel function. Using the properties
of Bessel functions, it is now easy to derive (\ref{ffins}).}
\beq
\int {d^d v \over (2 \pi)^d}  \,
{ e^{i v u} \over (v^2+{\tilde p}^2)^{{d \over 2}+2}} \; v_a v_b
\rightarrow_{u \rightarrow 0} {c \over 2 \p^2} \delta_{ab} \, .
\label{ffins}
\eeq
The relative factor $1/2$ 
between (\ref{ffin}) and (\ref{ffins}) is important to correctly 
reproduce the field theory results, as we will see later.
This support the usefulness of the closed string point of view
to understand UV/IR mixing effects, since the generalization of
the open Wilson line operators to include the adjoint scalars
(\ref{Ws}) is not dictated by gauge invariance.

\section{Strings on twisted circles}

In this and the following sections we will explore whether the 
remarkable connection between UV/IR mixing and closed strings found for 
D-branes at $\mathbf{C}^3/\mathbf{Z}_N$ orbifolds extends to more general 
cases. We will consider D-branes in Type II string theory compactified on 
twisted circle backgrounds. These are $\mathbf{R}^9 \times \mathbf{S}^1$ 
spacetimes where shifts along the circle are combined with rotations on 
several 2-planes \cite{Russo:2001na}. We will focus on the case 
where the rotations act on three planes and coincide with those 
that define a $\mathbf{C}^3/\mathbf{Z}_N$ orbifold
\beq
(y,z^l) \rightarrow (y +2 \pi R, e^{2 \pi i{b_l \over N}} z^l) \, ,
\label{tc}
\eeq 
where $y$ denotes the coordinate along the circle, $z^l$ the complex 
coordinates of $\mathbf{C}^3$ and $b_l$ are the twist parameters of 
the previous section. For $R \rightarrow \infty$ we recover
Type II string theory on $\mathbf{R}^{10}$. For $R=0$ the Type IIA(B)
twisted circle compactification reduces, after T-duality, to the associated 
$\mathbf{C}^3/\mathbf{Z}_N \otimes \mathbf{R}^4$ Type IIB(A) orbifold
model \cite{Takayanagi:2001jj}. 

The spectrum of closed strings on these backgrounds was derived in
\cite{Russo:2001na}. World-sheet bosons and fermions with indices transverse 
to the three 2-planes behave as in an ordinary circle compactification,
except for a different quantization condition on $p_y$.
Invariance of the wavefunctions under (\ref{tc}) requires 
\beq
p_y= {1 \over R}\left( m - \sum_l {b_l J_l \over N} \right) 
\;\; , \;\;\;\;\;\; m \in \mathbf{Z} \, ,
\label{py}
\eeq
where $J_l$ are the angular momenta on the three 2-planes.
The identification (\ref{tc}) links the winding number along the 
circle, $w$, with the boundary conditions for the world-sheet fields 
with indices along the 2-planes. For a given $w$, and for
the restricted set of examples we are considering, the left and 
right moving modes of these fields behave as those in the $k^{th}$ 
twisted sector of the associated $\mathbf{C}^3/\mathbf{Z}_N$ orbifold, 
with $k=w ({\rm mod} N)$. It is useful to define 
\beq
N'_k=N_k-c_k \; , \;\;\;\;\;\; {\bar N}'_k={\bar N}_k-{\bar c}_k \, , 
\eeq
with $N_k$ (${\bar N}_k$) the oscilator contribution
to $L_0$ (${\bar L}_0$) and $c_k$ (${\bar c}_k$) the zero-point energy of 
the left (right) modes in the NS or R $k^{th}$ twisted sector of the 
orbifold model. The mass spectrum of closed strings in the
twisted circle background (\ref{tc}) is then given by
\beq
M^2= {2 \over \alpha'} (N'_k+{\bar N}'_k) + {1 \over R^2}  
\left( m - \sum_l {b_l J_l \over N} \right)^2 + 
{w^2 R^2 \over \alpha'^2} \, ,
\label{spectrum}
\eeq
subject to the level matching constrain
\beq
N'_k-{\bar N}'_k= 
w \left( m - \sum_l {b_l J_l \over N} \right) \, .
\label{lm}
\eeq

$\mathbf{C}^3/\mathbf{Z}_N$ orbifolds generically have tachyons in the 
twisted sectors ($k \neq 0$). Therefore the associated twisted circle 
backgrounds may also have tachyons for appropriate radius, provided that
$w \neq 0({\rm mod}N)$.
Since the second term in (\ref{spectrum}) always gives a positive
contribution to the mass, let us consider the case it vanishes.
This implies: {\it i)} the level matching condition reduces to that of 
the related $\mathbf{C}^3/\mathbf{Z}_N \otimes \mathbf{R}^4$ orbifold model; 
{\it ii)} $\sum_l {b_l J_l \over N} \in \mathbf{Z}$, which imposses 
the orbifold projection.
Under these conditions, the first term in (\ref{spectrum})  
reproduces the masses of the orbifold $k^{th}$ twisted sector. 
The lowest mode in each winding sector of the twisted circle 
compactification is contained among this set. 
When $m_k^2<0$, with $m_k$ the mass of 
the lowest mode in the $k^{th}$ twisted sector of the associated 
orbifold \footnote{See \cite{Armoni:2003va} 
for the expression of 
$m_k^2$ in terms of the twist parameters $b_l$.},
the twisted circle compactification will contain tachyons for
radii
\beq
{R \over \sqrt{\alpha'}} < k^{-1} \sqrt{\alpha' |m^2_k|}\, ,
\label{thres}
\eeq
where we have taken $w=k$ in order to minimize the winding energy.

We will be interested in studying field theories arising on
noncommutative D-branes in these backgrounds. 
We will consider Type IIB D3-branes wrapped along the twisted circle 
and Type IIA D2-branes transverse to the twisted circle 
\cite{Dudas:2001ux,Takayanagi:2001aj}.
The intrinsic scale of the D3-brane theory is set by the Kaluza-Klein 
masses $m \sim 1/R$. Hence the
natural field theory limit for the D3-branes consist in sending  
$\alpha'\rightarrow 0$ keeping $R$ fixed. In this regime the 
winding energy dominates (\ref{spectrum}) and, 
as can be seen from (\ref{thres}), there are no tachyonic modes. 
Contrary, the natural scale of the D2-brane theory is $R/\alpha'$, 
governing the mass of strings with both ends on 
the brane and non-zero winding along the circle. The interesting field 
theory limit in this case is achieved by keeping $R/\alpha'$ fixed
as $\alpha'\rightarrow 0$. From (\ref{thres}) we observe
that this corresponds to the regime with closed string tachyons.
Therefore we obtain an optimal setup to explore the relation
between the leading UV/IR mixing effects and properties of the
closed string spectrum. Notice however that the noncommutative field 
theory is not able to resolve the value of $R$ for which tachyonic modes 
appear, since any $R \sim \sqrt{\alpha'}$ maps to $R=0$ in the field theory
limit appropriate for the D3-brane and 
$R/\alpha'=\infty$ for the D2-brane. 

\section{Wrapped D3-branes}

In this section it will be convenient to view the previous twisted circle 
backgrounds as freely acting orbifolds of 
$\mathbf{R}^9 \times \mathbf{S}^1_{R'}$, with $R'=N R$. We denote the 
identification (\ref{tc}) by $t$; it satisfies $t^N=1$. 

We place $n$ D3-branes on $\mathbf{R}^9 \times \mathbf{S}^1_{R'}$ wrapped 
along the circle and localized at $z^l=0$. The gauge theory on the D3-branes
is ${\cal N}=4$ $U(n)$. After performing the orbifold projection,
only modes whose wavefunction is left invariant survive
\beq
Z=\gamma_t \,( t \, Z )\, \gamma_t^{-1} \, ,
\label{orCP}
\eeq
where $Z$ is any of the ${\cal N}=4$ fields. The matrix
$\gamma_t$ represents the action of the orbifold on the Chan-Paton 
indices. Since it must satisfy $\gamma_t^N=1$, it can be taken
to be diagonal with eigenvalues $e^{2 \pi i j/N}$ appearing 
with multiplicity $n_j$ ($\sum n_j=n$), as for
$\mathbf{C}^3/\mathbf{Z}_N$ orbifolds. Let us decompose the field 
$Z$ in components $Z_{ij}$, transforming as $(\fund_i,\antifund_j)$ 
under the subgroup $\otimes U(n_j)$ which commutes with $\gamma_t$. 
The condition (\ref{orCP}) translates into a restriction on the allowed 
momenta $p_y$ analogous to (\ref{py})
\beq
p_y= {1 \over R}\left( m - {i-j + \sum_l b_l J_l \over N} \right) 
\;\; , \;\;\;\;\;\; m \in \mathbf{Z} \, .
\label{pD}
\eeq
 
\begin{table}
\begin{displaymath}
\begin{array}{|c|c|c|}
\hline
A^{ij}_{\mu \nu} & \psi^{ij}_\alpha & \phi_l^{ij} \\
\hline
{1 \over R}\big( m - {i-j \over  N} \big) &
{1 \over R}\big( m - {i-j + a_\alpha \over  N} \big) &
{1 \over R}\big( m - {i-j + b_l\over  N} \big) \\
\hline
\end{array}
\end{displaymath}
\caption{Field content of the theory and associated 
$p_y$. $A^{ij}_{\mu \nu}$ are vector fields,
$\psi^{ij}_\alpha$ ($\alpha=1,..,4$) Weyl fermions and 
$\phi_l^{ij}$ ($l=1,2,3$) complex scalars. $a_\alpha$ are the 
$\mathbf{C}^3/\mathbf{Z}_N$ twist parameters for
$SO(6)$ spinors, as in section 2.}
\end{table}

\noindent
Hence the theory on the wrapped D3-branes has the degrees of freedom of 
${\cal N}=4$ $U(n)$ compactified on a circle of radius $R$, with 
both supersymmetry and gauge symmetry broken by the different 
fractional momenta along the circle; see Table 1.

We turn on now a constant B-field along the two non-compact spatial
directions of the D3-brane, rendering its world-volume noncommutative.
Since the theory behaves at high energies as ${\cal N}=4$ Yang-Mills, 
there will be no UV divergence and thus no associated IR divergence. 
But given that supersymmetry is broken at the scale $1/NR$, we expect 
finite UV/IR mixing effects \cite{Landsteiner:2001ky}. 
From now on we will take the point of view of the dimensionally 
reduced theory and will focus on the zero-modes. The only 
fields with zero-modes on the circle are: $A_\mu^{ii}$, 
the gauge fields associated to $\otimes U(n_i)$, 
$\psi_\alpha^{ii+a_\alpha}$ and $\phi_l^{ii+b_l}$. This is the same set 
of fields that would survive the associated $\mathbf{C}^3/\mathbf{Z}_N$ 
orbifold projection. A complete treatment should of course include all 
fields and their Kaluza-Klein modes. The reason to focus on zero-modes is 
that the leading UV/IR mixing terms are simpler, and we expected them to 
have an stronger effect, on the massless 3d fields.

As we have already mentioned, the nonplanar 2-point functions 
project on $U(1)$ degrees of freedom and mix different group labels 
(see Fig. 2). We normalize the Lie algebra generators of each $U(n_i)$ 
such that ${\rm Tr} \, t_r t_s= \delta_{rs}$, with $t_0={1 \over \sqrt{n_i}} 
\mathbf{1}$. We take the external legs of the nonplanar 2-point functions 
to correspond to ${\rm Tr}A_\mu^{ii}=A_\mu^{ii\, (0)} \sqrt{n_i}$,
and analogously for adjoint scalars \footnote{We will discard backgrounds 
with some $a_\alpha=0$, since the associated orbifold and twisted circle 
backgrounds never contain closed string tachyons. Hence we will not
encounter adjoint spinors
with zero-modes.}. It is easy to see that with this choice of
normalization for the external legs, the nonplanar 2-point functions are
independent of the ranks $n_i$. We can then proceed
with the evaluation as if our gauge group would be $U(1)^N$.
The 2-point functions for the zero-modes in that case are calculated 
in detail, at leading order, in an appendix.

Let us introduce some convenient notation. We define 
$\Pi_{ij}^{(1)}$ by $\Pi_{ij}^{\mu \nu}= {\p^\mu \p^\nu \over \p^2}
\, \Pi_{ij}^{(1)}$, where $\mu=1,2,3$ are the three noncompact
directions, and $\Pi_{ij}^{(2)}=\Pi_{ij}^{yy}$. All other components of 
the polarization tensor for zero-modes are zero. The matrix 
$\Pi_{ij}^{(3)}$ will refer to the self-energies of adjoint scalar 
zero-modes when present (this requires that some $b_l=0$),
$\Sigma_{ij}^{ab}=\delta^{ab} \, \Pi_{ij}^{(3)}$. Using this compact 
notation and (\ref{apinp1})-(\ref{apinp3}), the nonplanar
2-point functions are given by
\beq
\Pi^{(M)}_{ij}= {g^2 \over 2 \pi^2 \p} \, c_{ij} \sum_\rho (-1)^{F_\rho} 
\! \int_0^\infty d l \, N_{{\bar \nu}_\rho}(Rl) \, f^{(M)}({\tilde p}l)  \, .
\label{pift}
\eeq
The index $\rho$ runs over all degrees of freedom transforming in the
$(\fund_i,\antifund_j)$ and $(\fund_j,\antifund_i)$ representations, 
and $F_\rho$ is their fermion number. The exchange of an adjoint field
in the loop contributes twice that of a bifundamental, thus $c_{ii}=1$ 
while $c_{ij}=1/2$ for $i \neq j$. The function $N_{{\bar \nu}_\rho}$ 
is given in (\ref{N}). It represents the generalization of the Bose-Einstein 
and Fermi-Dirac distributions to fields with 
general fractional momenta along a circle,
$p_y=(m-{\bar \nu}_\rho)/R$. The functions $f^{(M)}$ are
\beq
f^{(M)} (x) = 
\sin x +b^{(M)} x \cos x \, , 
\eeq
with $b^{(1)}=-1$, $b^{(2)}=1$ and $b^{(3)}=0$. 

The integrals in (\ref{pift}) can be easily evaluated 
recalling (\ref{dis}) and performing a Poisson resummation
\beq
\int_0^\infty dl N_{{\bar \nu}_\rho}(R l) \sin \p l  =  \pi T \sum_{k=0}^{N-1} 
\cos \big( 2 \pi {\bar \nu}_\rho k \big) \, N_{\nu_k} (T \p) \, , 
\label{poisson}
\eeq
where we have defined $\nu_k=k/N$ and $T=1/2 \pi R N$. 
$\int dl N_{{\bar \nu}_\rho} l \cos \p l$ can be directly obtained
from the above expression, using that $l \cos \p l={\partial \sin \p l  
\over \partial \p}$. Substituting these results, together
with the explicit expression of $N_{\nu_k}$ (\ref{N}), we arrive at 
\beq
\Pi^{(M)}_{ij}= g^2 T^2  \;
\sum_{k=0}^{N-1} M_{ij;k} \;  F_k^{(M)}(T {\tilde p}) \, ,
\label{polt}
\eeq
with 
\beq 
F^{(M)} (x) =  {1 \over \cosh 2 \pi x -\cos 2 \pi \nu_k}
 \left[\, {\sinh 2 \pi x \over 2 \pi x }+ b^{(M)} \,
{1-\cos 2 \pi \nu_k \cosh 2\pi x \over \cosh 2 \pi x -\cos 2 \pi \nu_k}
\, \right]    \, ,
\label{sums}
\eeq
and $M_{ij;k} = c_{ij} \sum_\rho (-1)^{F_\rho} \cos 2 \pi {\bar \nu}_\rho k$. 
From the data of Table 1, an straightforward 
calculation gives 
\beq
M_{ij;k} = \epsilon_k \, \cos \big( 2 \pi (i-j) \nu_k \big) \, .
\label{mij}
\eeq
Remarkably the quantities $\epsilon_k$ coincide with those 
resulting from the diagonalization of the polarization tensor for 
the $\mathbf{C}^3 / \mathbf{Z}_N$ quiver theories analyzed in section 2, 
eq. (\ref{ep}).
However the matrices $\Pi_{ij}^{(M)}$ are not yet in a diagonal form. 
Notice that their $ij$ entry depends only on 
$(i-j)\, {\rm mod} N$. Any matrix with this
property can be diagonalized by a set of orthogonal vectors 
$(e_k)_j={1 \over \sqrt{N}} e^{2 \pi i j k/N}$. In the new basis (\ref{polt})
reads  
\beq
\Pi^{(M)}_k=  g^2 T^2 \, N \; \epsilon_k \; 
F_k^{(M)} (T \p) \, .
\label{pid}
\eeq

The sign of the 2-point functions is governed by $\epsilon_k$, as it 
was the case for the $\mathbf{C}^3 / \mathbf{Z}_N$ quiver theories,
because the functions $F_k^ {(M)}$ are always positive. 
This can easily be shown by direct inspection of (\ref{sums}).
For $k \neq 0$, $F_k^{(M)}$ is finite as $\p \rightarrow 0$. 
The case $k=0$ is special. $F_0^{(M)}$ diverges for small $\p$ since
this sector contains the IR divergences of the noncompact 4d theory.
However, $\epsilon_0=0$ because the uncompatified limit is supersymmetric. 
As $T\rightarrow \infty$, $F_k^{(M)}$ tends to $1/2 \pi T \p$.
In this limit (\ref{pid}) reproduces the $1/\p \,$ IR divergences of 
the $\mathbf{C}^3 / \mathbf{Z}_N$ quiver theory on a D2-brane. 
Therefore at some point of the flow to small radius
(\ref{pid}) will dominate the dispersion relations for low momentum
modes, and noncommutative instabilities will appear.

\begin{figure}
\centering
\epsfxsize=2.5in
\hspace*{0in}\vspace*{0in}
\epsffile{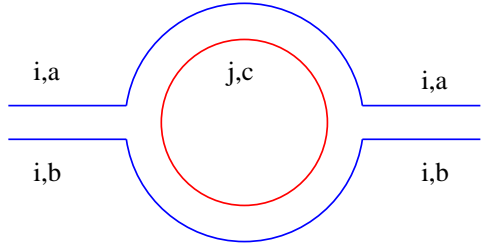}
\caption{\small Planar contribution to the 2-point function of adjoint 
fields. $i,j$ are group label indices and $a,b,c$ gauge group indices.}
\end{figure}

The analysis of the critical radius at which some modes become unstable
is quite involved. It strongly depends on which fields we are considering, 
vector or scalar, on the sector $k$ and on the particular brane
configuration, characterized by the ranks $n_i$. We will not go through 
this complicated phenomenology, because what interests us is the structure 
behind these effects. However for completion, we include below the
leading planar contributions to the 2-point functions. 

We reduce again to the zero-modes of gauge bosons and adjoint scalars. 
As can be seen from Fig.3, in this 
case there is no mixing among the different group labels. The planar
correction for the 3-dimensional gauge bosons is absent; for the others
fields we have (see (\ref{apip}))
\beq
{\hat \Pi}^{(2)}_{ii}= 2 {\hat \Pi}^{(3)}_{ii}=- {g^2 \over \pi^2} \, 
\sum_\rho c_\rho \, (-1)^{F_\rho} \! 
\int_0^\infty d l \, N_{{\bar \nu}_\rho}(Rl) 
\, l   \, .
\label{pi}
\eeq
The summation in $\rho$ is over the degrees of
freedom transforming in the $(\fund_i,\antifund_j)$ and 
$(\fund_j,\antifund_i)$, for any $j$. $c_\rho=1$ for
a field in the adjoint representation, while $c_\rho=1/2$ for
a bifundamental.
The external legs will be normalized now such that ${\hat \Pi}_{ii}^{(2)}$
couples to ${\rm Tr}(A_y^{ii})^2$, which is appropriate for a planar graph,
and analogously for ${\hat \Pi}_{ii}^{(3)}$.
With this normalization, the 2-point functions will
include an explicit dependence on the ranks $n_j$. We obtain
\beq
{\hat \Pi}^{(2)}_{ii}= -{2 g^2 T^2 \over \pi^2 } \;
\sum_{k=0}^{N-1} \epsilon_k  \left( \sum_{j=1}^N \, n_j\,\cos 
\big(2 \pi (i-j) \nu_k \big) \right) Z_2[\nu_k] \, ,
\label{planar}
\eeq 
with $Z_m[x]$ a generalized Riemann Zeta function.

It is worth stressing that the result of evaluating the integrals 
in (\ref{pift}) was given again in terms  of functions 
$N_{\nu_k}$, with their argument changed from $Rl$ to $T \p$.
A similar observation, in the context of noncommutative 
thermal field theories, was done in \cite{Fischler:2000fv}.
Since $T \sim 1/R$, they suggested that there must be some
hidden winding states in the noncommutative theory.
In \cite{Arcioni:2000bz} it was argued that these mysterious states
were closed strings winding around the temperature circle
of a parent string theory. They noticed that an infinite
set of closed strings must play a role, very much as in
ordinary field theory limits of string theory. This raised the question 
about the usefulness of the closed string picture. 
In the next section we will present a very explicit realization  
of these ideas along the lines of section 2. In particular
we will show that the closed string picture is indeed useful to 
understand the structure of both (\ref{pid}) and (\ref{planar}).

\section{Closed string exchange}

Let us analyze what type of Wilson line operators couple to 
closed strings in twisted circle backgrounds. It is useful to consider 
first D3-branes wrapped along an ordinary circle of radius $R'=NR$.
Closed strings couple to the straight open Wilson line operators (\ref{Ws})
on noncommutative D-branes. These operators depend on the 
components of the gauge field along the noncommutative directions and on 
adjoint scalars, associated to the position of the branes in transverse 
dimensions. However this must be generalized when one direction on the
D-brane is compact. T-dualizing, the D3-brane is converted
to a D2-brane localized on the circle. The zero-mode of $A_y$ describes 
the positions of branes on the T-dual circle and thus plays a similar role 
to the adjoint scalars. Hence in order to be consistent with T-duality, 
we should include a dependence on $A_y$ in the exponent of the Wilson line.

The explicit construction of the open Wilson line operators that are
gauge invariant under the 4d gauge transformations is very involved.
However we will not need to know their precise expression. We are only 
interested on the coupling of closed strings to the zero-modes of gauge 
fields and scalars, since this will be enough to reproduce the results 
of the past section. With this restriction, we propose the following 
modification of (\ref{Ws})
\beq
{\widetilde W}_{p_y=0}  = 2 \pi R' \, {\rm Tr} \int d^3 x \; P_\ast 
\left(e^{i\, g \int_0^1 d \sigma 
\, \big( {\tilde p}^\mu A_{\mu}(x(\sigma))+
v_a \phi_a(x(\sigma))+ 2 \pi R' s  A_y (x(\sigma))\big)}\right) 
\ast e^{i p x} \,+ \, ...  \, ,
\label{wsw}
\eeq
where $A_\mu(x)={1 \over 2 \pi R'} \int dy A_\mu(x,y)$ and the same 
for $A_y$ and $\phi_a$; we have denoted $x(\sigma)=x+\p \sigma$. 
The dots indicate terms depending on modes with $p_y\neq 0$, necessary 
to make ${\widetilde W}$ 4d gauge invariant. The term written explicitly 
above is however gauge invariant under 3d gauge transformations. 
As in section 2, $v=2 \pi \alpha' p_{\perp}$, with $p_\perp$ the momentum 
on the transverse directions of the associated closed string mode. 
$s \in \mathbf{Z}$ is its winding number. Notice that $2 \pi R's= 
2 \pi \alpha' p_{\bar v}$, with 
$p_{\bar v}=s R'/ \alpha'$ the momentum on the dual T-circle.
The factor $2 \pi R'$ in (\ref{wsw}) 
comes from the trivial integration along the circle.  

The interpretation of the twisted circle backgrounds (\ref{tc}) as
freely acting orbifolds of $\mathbf{R}^9 \otimes \mathbf{S}^1_{R'}$,
divides the closed string spectrum (\ref{spectrum}) in twisted and
untwisted sectors. The untwisted sector is composed of modes
with $w=0({\rm mod}N)$. The $k^{th}$ twisted sector ($k=1,..,N-1$) 
contains modes with $w=k({\rm mod}N)$. 
These modes will couple to linear combinations of field theory operators 
determined by $\gamma_t$, the matrix that represents the action of the 
orbifold on the Chan-Paton indices \cite{Douglas:1996sw}. 
Since $\gamma_t$ for the twisted circle compactifications coincides 
with that of $\mathbf{C}^3/\mathbf{Z}_N$ orbifolds, we should consider 
the same linear combinations as in that case. This suggests that an scalar 
closed string mode with $w=k({\rm mod}N)$ will couple to
\beq
W^{(k)}= {1 \over \sqrt{N}} \sum e^{2 \pi i j \nu_k} 
{\widetilde W}^{(j)}(\nu_k) \, ,
\eeq
where ${\widetilde W}^{(j)}(\nu_k)$ is
the operator (\ref{wsw}) with the fields belonging to the
$j^{th}$ gauge group factor and $s$ substituted by
$(\nu_k+s)=w N^{-1}$.

In the noncommutative field theory limit \cite{Seiberg:1999vs}, taken at 
fixed radius, the on-shell condition a closed string mode
$\varphi$ with $w=k({\rm mod}N)$ is
\beq
{\tilde p}^2+y^2+ (2 \pi R')^2 (\nu_k+s)^2+ 8 \pi^2 \alpha'
(N'_k+{\bar N}'_k)=0 \, ,
\label{spsw}
\eeq
where we have used (\ref{spectrum}) and neglected terms suppressed by 
two powers of $\alpha'$. 
We have now all the ingredients to derive the contribution to the field 
theory effective action from a closed string exchange, following the
analysis of section 2. Generalizing
(\ref{staction}) and (\ref{actsc}) to the present case, we obtain from the 
emission, propagation and posterior absorption of $\varphi$ by the brane 
\beq
\Delta S = {h_\varphi \over 2 \pi R'} \; 
\sum_{s=-\infty}^{\infty}  \int {d^3 p \over (2 \pi)^3} \, 
{d^d v \over (2 \pi)^d} \; { W^{(k)}(P) \; W^{(N-k)}(-P)   \over 
\Big(\,v^2+{\tilde p}^2 + (2 \pi R')^2(\nu_k+s)^2 \, \Big)^{{d \over 2}+2}} 
\, e^{iyu} \, ,
\label{saction}
\eeq
where $P\equiv(\p,v,\nu_k+s)$, $d$ is the number of transversal 
directions where the closed string can propagate. $h_\varphi$ depends on 
the disk amplitude with one insertion of $\varphi$
and on the last term in (\ref{spsw}). 

All scalar closed string modes contribute to the field theory
effective action as in (\ref{saction}), only differing in the value of 
$h_{\varphi}$. Since $W^{(k)}$ (\ref{spsw}) carries no momentum along
the circle, it can only couple to closed string modes with $p_y=0$. 
In section 3 we saw that for these modes, the last term
in (\ref{spsw}) reproduces the spectrum in the $k^{th}$ twisted
sector of the related $\mathbf{C}^3/\mathbf{Z}_N$ orbifold.
We will assume that the disk amplitudes with 
one insertion of these modes and no open string insertions also coincide 
for orbifolds and twisted circles backgrounds.  
With this hypothesis, after taking into account the effect
of the whole closed string tower, $h_\varphi$ should be promoted to the same 
quantities that govern the effective action of D3-branes in 
$\mathbf{C}^3/\mathbf{Z}_N$ orbifolds (\ref{actsc}): 
$\epsilon_k / 2 \pi^2 c$, with $c$ given in (\ref{ffin}) and (\ref{ffins}).

We are ready to derive the contributions to the 2-point functions predicted
by the closed exchange technique. We study first the non-planar contributions.
They are obtained by expanding the Wilson line operators to first order
in the fields. The linear combinations that 
diagonalize $\Pi_{ij}^{(M)}$ appear in a natural way: 
$B_M^{(k)}={1 \over \sqrt{N}} \sum e^{2 \pi i j \nu_k} {\rm Tr} Z_M^{(j)}$,
with $Z_M^{(j)}=(A_\mu,A_y,\phi_a)^{jj}$. In order to compare with 
previous results, it is important to take into account that the fields
in the exponent of (\ref{wsw}) are $Z^{(j)}(x)={1 \over 2 \pi R} 
Z^{(j)}_{n=0}(x)$, with $Z^{(j)}_{n=0}$ the zero-modes for which
the 2-point functions were calculated in the past section. 
From (\ref{saction}) we obtain then an
expression for $\Pi^{(M)}_k$ analogous to (\ref{pid}), with the functions
$F_k^{(M)}$ replaced by
\beqa
F_k^{(1)st} & = & 4 R'^2 \, \sum_{s=-\infty}^{\infty} 
{\p^2 \over \Big( \, {\tilde p}^2 + (2 \pi R')^2(\nu_k +s)^2 \, \Big)^2} 
\label{pi1} \, ,\\
F_k^{(2)st} & = & 4 R'^2 \, \sum_{s=-\infty}^{\infty} 
{(2 \pi R')^2 (\nu_k+s)^2 
\over \Big( \, {\tilde p}^2 + (2 \pi R')^2(\nu_k +s)^2 \, \Big)^2} \, , \\
F_k^{(3)st} & = & 2 R'^2 \,  \sum_{s=-\infty}^{\infty}
{1 \over {\tilde p}^2 + (2 \pi R')^2(\nu_k+s)^2} \, . \label{pi3} 
\eeqa
For the origin of the relative factor $1/2$ in $F_k^{(3)st}$
see (\ref{ffins}). Using (\ref{dis})-(\ref{N}), it is easy to show that 
$F_k^{(M)st}=F_k^{(M)}$.
$\Pi_k^{a\mu}=\Pi_k^{ay}=0$ because the associated integrand is an odd 
function of $y$. $\Pi_k^{y\mu}=0$ is obtained by summing the contributions 
from the $k^{th}$ and $(N-k)^{th}$ sectors. 

To derive the planar contribution to the 2-point functions from 
(\ref{saction}), we expand one Wilson line operator to second order in
the fields while retaining from the second one only the leading term. 
Due to the structure of (\ref{saction}),
we get only contributions diagonal in the 
group label indices, as was shown to be the case in the field theory 
analysis (see Fig. 3). The leading term in the expansion of $W^{(k)}$ is
proportional to ${1 \over \sqrt{N}} \sum e^{2 \pi i j \nu_k} n_j \, 
\delta^{(3)}(p)$. Since the closed string propagator is finite as 
${\tilde p}\rightarrow 0$ while the gauge field $A_\mu$ appears always 
multiplied by corresponding ${\tilde p}^\mu$ factors, the result for 
${\hat \Pi}^{\mu \nu}_{ii}$ is zero. For the other components we get exactly
(\ref{planar}).

The contribution to the 1-loop effective action from the vacuum
diagrams, $\Gamma$, can be evaluated using that 
\beq
{\partial \Gamma \over \partial \beta}= - {V_3 \over 2}
\, \sum_\rho (-1)^{F_\rho} \int 
{d^3 l \over (2 \pi)^3} N_{{\bar \nu}_\rho} \, l \, ,
\eeq
where $\beta= 2 \pi R$ and $V_3$ represents the volume of the three 
noncompact dimensions. The integration is over the euclidean 
3-momentum and $\rho$ runs over all degrees of freedom of the theory. 
We obtain
\beq
\Gamma = {V_4 \, T^4 \over \pi^2} \, \sum_{i,j} \, n_i \, n_j\, 
\sum_k \epsilon_k \, \cos \big(2 \pi (i-j) \nu_k \big)  \, Z_4[\nu_k] \, ,
\eeq
with $V_4=2 \pi R \, V_3$. Retaining only the first term in the 
expansion of both Wilson line operators, one can check that 
(\ref{saction}) also reproduces 
$\Gamma$. It was argued in \cite{Armoni:2003va} that $\Gamma$ for the 
$\mathbf{C}^3/\mathbf{Z}_N$ quiver theories of section 2, could be 
recovered from (\ref{invaction}) in an analogous way. However this implied
identifying ${1 \over \p}|_{\p=0}$ with an UV cutoff $\Lambda$, 
and introduced certain ambiguity. This problem does not arise
in the theories treated here, since they are UV finite.

Let us summarize our results. The closed string picture
captures both the leading planar and non-planar terms which will give 
rise in the limit $T \rightarrow \infty$ to the UV and IR divergences 
linked to the absence of supersymmetry. Moreover, it offers an 
explanation for why the same coefficients $\epsilon_k$ appear in
$\mathbf{C}^3/\mathbf{Z}_N$ orbifolds and twisted circle backgrounds.
Twisted circle backgrounds, for sufficiently large radius, 
cure the problem of closed string tachyons by shifting the spectrum 
with a positive winding energy. However this affects equally 
NSNS and RR sectors. Since NSNS and RR sectors contribute with 
opposite sign to (\ref{saction}), this shift is not perceived by 
the field theory. The sign of the leading UV/IR mixing effects is 
still given by that of the gap between the lowest NSNS and RR modes.

The nonplanar functions that we calculated in this and the past 
section are regular in $\p$. From the string point of view, this
crucially depends on being in a regime where winding energy
dominates the closed string spectrum. Instabilities arising from
finite UV/IR effects can be expected to have a qualitatively different 
nature from those associated to IR divergences. It is then an open 
question whether the one to one correspondence between tachyons
and noncommutative instabilities found for $\mathbf{C}^3/\mathbf{Z}_N$ 
orbifolds, might hold in more general IR divergent cases. The fact that 
the nonplanar polarization tensor (\ref{pid}) develops a pole-like 
behavior at $R=0$, value which maps to the string regime with tachyons,
supports this possibility. 
The twisted circle backgrounds provide us with another example of a
field theory related to an unstable string regime: that on the 
world-volume of a D2-brane placed transverse 
to the circle. In the next section we will analyze the leading
UV/IR mixing effects in this theory.

\section{Transversal D2-branes}

We consider $n$ Type IIA D2-branes extending in the 
directions transverse to (\ref{tc}) and localized at $z^l=y=0$. 
It is convenient to picture the brane configuration on the $\mathbf{R}^{10}$
covering space. It corresponds to an infinite array of D2-brane
sets placed at intervals $2 \pi R$ on the $y$ direction.
Let us denote open strings stretching between the $i^{th}$ and
$j^{th}$ sets by $(i,j)$. The action (\ref{tc}) relates $(i,j)$ 
strings with $(i+1,j+1)$, but does not impose any projection
on each of them separately. Thus the spectrum on the branes will be 
maximally supersymmetric \cite{Dudas:2001ux}. We are interested in 
taking the field theory limit such that the masses of the winding 
strings are keep finite, {\it i.e.} $\alpha'\rightarrow 0$ with 
$R/\alpha'=r$ fixed. The field content on the branes will
consist of an infinite set of multiplets $\Psi_w$ 
with masses $m_w= w r$. Each multiplet $\Psi_w$
transforms in the adjoint of $U(n)$ and contains 3 complex scalars 
$\phi_{l,w}$, 4 Weyl fermions $\psi_{\alpha,w}$
and, for $w\neq 0$, a massive vector $A_{\mu,w}$, while for
$w=0$ a massless vector plus a real scalar $A_{\mu,0}$, $\varphi_0$.
The scalars $\phi_{l,0}$, $\varphi_0$ are associated to fluctuations 
of the branes in the $z^l$ and $y$ directions. The previous field theory 
limit corresponds to the string regime with tachyonic modes, as can be 
seen from (\ref{thres}).

In this section we will focus on the evaluation of the polarization tensor. 
We turn on a B-field on the two spatial directions of the D2-branes.
We will analyze first what the closed string picture predicts for the 
leading UV/IR mixing effects. As before, closed strings will couple to 
open Wilson line operators. The explicit construction of these operators 
is again involved. For our purposes, it will be enough to know few terms
in their expansion 
\beq
W^{(w)}(p)= \delta_{w,0} \, (2 \pi)^3 \delta^{(3)} (p) {\rm Tr} {\mathbf{1}}  
+ i g \p^\mu {\rm Tr} A_{\mu,w}(p)+... \, .
\label{Ww}
\eeq
Given that the identity operator carries no winding number, it can only
appear in the expansion of $W^{(0)}$.

Closed string modes that couple to (\ref{Ww}) have winding number 
$w$. Since the D2-branes break translational invariance along the $y$ 
direction, their momentum $p_y$ is a free parameter. There is however 
a restriction. D2-branes located at $z^l=0$ do not break rotational 
invariance on $C^3$, and thus their vacuum state carries no 
angular momentum. Neither do the fields $A_{\mu,w}$. Therefore (\ref{py}) 
implies $p_y=m/R$ with $m \in \mathbf{Z}$. 

In the noncommutative field theory limit that we are considering in this
section, the on-shell condition for a closed string mode $\varphi$ with 
$w=k({\rm mod}N)$ and $p_y=m/R$ reads
\beq
{\tilde p}^2+y^2+ {4 \pi^2 m^2 \over r^2} + 
8 \pi^2 \alpha'(N'_k+{\bar N}'_k)=0 \, ,
\label{spsw2}
\eeq
where again we have neglected terms suppressed by two $\alpha'$
powers. Using the same reasoning as in the past sections,
we obtain the following contribution to the transversal part of the
polarization tensor from a closed string exchange
\beq
\Pi^{\mu \nu}_w \sim h_\varphi \; r^{-1} g^2   {\p^\mu \p^\nu \over 
\Big(\,{\tilde p}^2 + {4 \pi^2 m^2 \over r^2} \, \Big)^2} \, .
\label{strexch}
\eeq
Notice that the dependence on $\p$ is that proper of a 4d theory.
This arises as follows. The definition of the closed 
string propagator includes now a factor $1/2 \pi R$, because the 
brane is transversal to the circle. In the $\alpha'\rightarrow 0$ 
limit that keeps finite the mass of winding strings, $R$ absorbs 
one $\alpha'$ factor and gives raise to the $r^{-1}$ factor in
(\ref{strexch}). As a result, the counting of 
$\alpha'$ powers is as for the D3-brane case.

The combination $2 \pi m/r$ in (\ref{strexch}) plays an analogous role 
to the winding energy in (\ref{saction}). There is however a crucial
difference; in the latter case, the winding energy could not be zero if 
$w \neq 0({\rm mod}N)$. Contrary, $m=0$ is an allowed value for any 
$w$. It is clear that $m=0$ will dominate the small $\p$ limit of 
the sum in (\ref{strexch}). Let us focus then on this value. 

The quantity $h_\varphi$ depends on the last term in (\ref{spsw2}). 
When $m=0$, the associated closed string mode has $p_y=0$.
As we have already explained, the last term in (\ref{spsw2}) reproduces 
then the spectrum of the $k=w ({\rm mod} N)$ twisted sector of the related
$\mathbf{C}^3 /\mathbf{Z}_N$ orbifold.
$h_\varphi$ depends also on a numerical factor, $D_\varphi^{tc}$,
which can be extracted from the disk amplitudes 
of $\varphi$ with boundary conditions on the D2-brane.
However we can not take now the brane in its vacuum state, since
this restricts us to $w=0$. Let us consider the disk amplitude with one 
insertion of $A_{\mu,w}$ on its boundary. This should reproduce 
$D_\varphi^{tc}$ times the second term in (\ref{Ww}). 
As in the past section, we will assume that $D_\varphi^{tc}$ coincides 
with the analogous factors from the orbifold case of section 2. 
With that hypothesis, and after taking into 
account the contribution from all closed string modes to 
(\ref{strexch}), $h_\varphi$ will be promoted to the same
quantities $\epsilon_k$ that appeared in the previous
sections. Hence, at small $\p$, the closed string picture suggests
the following result for the polarization tensor
\beq
\Pi^{\mu \nu}_w \sim \epsilon_k \; r^{-1} g^2   {\p^\mu \p^\nu \over 
{\tilde p}^4} \, .
\label{strexch2}
\eeq
with $k=w({\rm mod}N)$.

We will like to add a comment. Comparison with (\ref{Ws}), for $w=0$, 
implies that the expansion of $W^{(0)}$ at leading order in the fields
should include a term $\sim \alpha' p_y {\rm Tr} \, \varphi_0= {m \over r} 
{\rm Tr} \, \varphi_0$, where $\varphi_0$ is the massless scalar of
the $\Psi_0$ multiplet. For $W^{(w)}$, with $w \neq 0$, we expect that 
this is promoted to a term $\sim {m \over r}{\rm Tr} A_{L,w}$, 
where $A_{L,w}$ denotes the longitudinal component of the massive 
gauge field $A_{\mu,w}$. Using this and the closed string exchange
technique, we could try to derive a contribution to the longitudinal 
piece of the polarization tensor. However the string derivation 
predicts no contribution from $m=0$, and thus we have ignored it.
This is consistent with the field theory analysis below.

We will confirm now the proposal (\ref{strexch2})
from a direct field theory calculation. The appearance
of IR divergences could have surprised us, since the field content 
of the D2-brane theory is supersymmetric. Notice that 
the spectrum on D3-branes, both at orbifold singularities
and twisted circles, was not supersymmetric.
The coefficients $\epsilon_k$ were essentially counting there the number 
of bosonic minus fermionic degrees of freedom from the point of view of
the field theory. 
Contrary to those cases, supersymmetry is broken by interactions
on the D2-brane. Thus we need to know how the Feynman rules 
get modified.

Let us come back to image of the D2-brane configuration on the
covering space. We will denote as $Z_{(i,j)}$ a field arising from an 
string stretching between the $i^{th}$ and $j^{th}$ D2-brane sets.
Before imposing (\ref{tc}), our gauge theory is $U(\infty)$
broken to an infinite product of $U(n)$ factors by the separation of
the branes. A field $Z_{(i,j)}$ transforms in the fundamental 
representation of the
gauge group factor $i^{th}$ and antifundamental of the $j^{th}$.
Therefore the only terms that can appear in the lagrangian of the 
theory form closed paths in the direction $y$; for example
\footnote{These terms contain also derivatives. We are
being schematic for simplicity.}
\beq
{\rm Tr} \; {\bar Z}_{(i,j)} Z_{(j,i)} \; , \;\;\;
{\rm Tr} \; Z^1_{(i,j)}Z^2_{(j,k)}Z^3_{(k,i)} \, .
\eeq
All the kinetic terms are canonically normalized and the
three- and four-point vertices depend on a single coupling constant
$g$. 

Invariance under (\ref{tc}) implies
\beq
Z_{(i,j)}= \omega_Z^i \; Z_{(0,j-i)} \, ,
\label{fields}
\eeq
with $\omega_{\phi_l}=e^{2 \pi i {b_l \over N}}$, 
$\omega_{\psi_\alpha}=e^{2 \pi i {a_\alpha \over N}}$
and $\omega_{A_\mu}=1$. From now on we will take the fields 
$Z_{(0,w)}$ as reference and we will label them simply as $Z_w$.
They form the multiplets that we defined in the beginning of
the section as $\Psi_w$.

In order to obtain the polarization tensor for the vector fields
$A_{\mu,w}$, we have to evaluate the graphs in Fig. 2 and 3.
We need to know the propagators and vertices of the theory in terms 
of the fields $Z_w$. 
Since ${\bar Z}_{(i,j)} Z_{(j,i)}= \omega_Z^{-w} {\bar Z}_{-w} Z_w$,
with $w=i-j$, and the field $Z_{(i,j)}$ is canonically normalized,
the propagator for the field $Z_w$ will carry an additional factor
$\omega_Z^w$. For constructing the graphs in Fig. 2 and 3, two 
vertices are relevant
\beq
V_1= g_1 \; {\rm Tr} \; {\bar Z}_{w_1} A_{\mu,w} Z_{w_2} \;\; , \;\;\;\;\;
V_2=- g_2 \; {\rm Tr} \; {\bar Z}_{w_1} Z_{w_2} A_{\mu,w} \, ,
\eeq
with $w_1+w_2+w=0$ and $g_1$, $g_2$ the coupling constant
associated to each vertex. Using (\ref{fields}), it is easy to see 
that $g_1=g \, \omega_Z^{-w_2}$ and $g_2=g \, \omega_Z^{w_1}$. The 
different powers of $\omega_Z$ associated to the two vertices may 
result surprising. However we should notice that the two 
vertices come from different configurations in the covering
space, see Fig. 4.

\begin{figure}
\centering
\epsfxsize=4.2in
\hspace*{0in}\vspace*{0in}
\epsffile{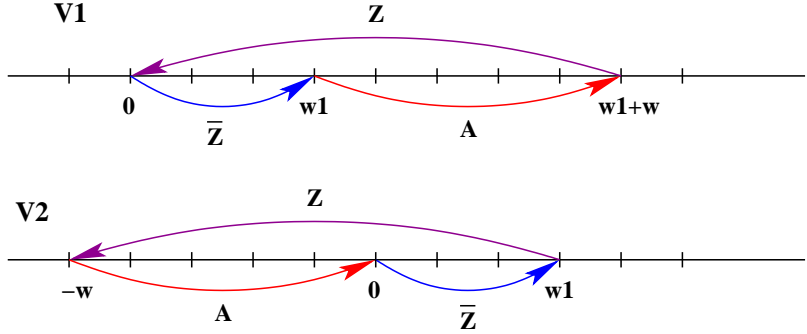}
\caption{\small Interaction vertices $V_1$ and $V_2$ in the covering space.}
\label{chain}
\end{figure}

The non-planar graph of Fig. 2 arises when the two vertices
are $V_i$, $V_j$ with $i \neq j$ ($i,j=1,2$). 
Using the rules explained above, we obtain the following
contribution to the non-planar polarization tensor from a field 
$Z$ running in the loop
\beq
{1 \over 2}\big( \omega_Z^{w}+\omega_Z^{-w} \big) \, \Pi^{\mu \nu}_Z \, ,
\label{phs}
\eeq 
where $\Pi^{\mu \nu}_Z$ would be the analogous result when 
$\omega_Z=1$. This simpler situation corresponds to D2-branes placed 
transversally to an ordinary circle, and is T-dual to D3-branes wrapping 
an ordinary circle of radius $r^{-1}$. The evaluation of 1-loop integrals 
is most convenient in the T-dual picture. It results in a piece which is 
$r$-independent and reproduces the uncompactified
$r^{-1} \rightarrow \infty$ result, and a part that is UV finite and $r$ 
dependent. The former one gives rise to an IR divergence
and dominates in the limit of small $\p$. Since this is the IR
divergence of a 4d theory, we have\footnote{In addition to the 
graphs in Fig. 2, there are 1-loop diagrams
with the insertion of a four-point vertex contributing to the 
nonplanar polarization tensor. The same considerations above
can be applied to them with similar results.}
$\Pi^{\mu \nu}_Z \sim r^{-1} g^2 \, {\p^\mu \p^\nu  \over \p^4}$, 
with $r^{-1} g^2$ the 4d coupling constant. Summing now to 
$Z=(\phi_l, \psi_\alpha, A_\mu)$ and having into account the phases
in (\ref{phs}), the parameters $\epsilon_k$
reappear and we recover (\ref{strexch2}). 

It is also interesting to evaluate the planar piece of the 
polarization tensor, Fig. 3. It arises when both vertices are $V_1$ 
or $V_2$. In these cases, the dependence on the phases $\omega_Z$ 
cancels. Since the spectrum of the theory is
supersymmetric, adding the contributions from all fields will give a
vanishing result. For the same reason the 1-loop vacuum 
energy of the theory is zero. Both results are predicted by the closed
string picture. Notice that the vacuum diagrams can only come from the
leading term in the expansion of $W^{(0)}$, and the planar 
part of the polarization tensor from higher terms in the 
expansion of $W^{(0)}$, which we have not written explicitly in (\ref{Ww}). 
Since $\epsilon_0=0$, we obtain no contribution.

\section{Conclusions}

The main motivation of this paper was to answer some of the open questions 
left in \cite{Armoni:2003va}, where a direct relation between noncommutative
instabilities and closed string tachyons was found for a family of gauge 
theories on D3-branes at $\mathbf{C^3}/\mathbf{Z_N}$ orbifolds. 
In this section we summarize our results.

The first question was whether the relation between noncommutative
instabilities and closed string tachyons holds in other examples. A useful
tool to study this problem was a proposal to derive UV/IR mixing 
effects purely in the language of closed strings. This point of
view supported the distinction of two separate pieces in the leading 
UV/IR mixing effects: their functional dependence on $\p$ and 
their sign. The sign determines if the leading UV/IR mixing effects 
will tend to destabilize or not the theory. The functional 
dependence on $\p$ can be finite or divergent in the IR depending if
supersymmetry is or not restored at some high scale.

The closed string derivation suggested that the sign of the leading UV/IR 
mixing effects is linked to the misalignment between the NSNS and RR 
towers of the parent string theory. In order to check this proposal, 
we have analyzed in this paper twisted circle compactifications of
Type II string theory. The NSNS and RR towers are shifted in the
same amount by a positive winding energy. Tachyonic modes
appear only for twisted circles of sufficiently small radius.
We have studied Type IIB D3-branes wrapped on the twisted circle
and Type IIA D2-branes transversal to it. We showed that in both cases
the UV/IR mixing signs are governed by the $($mass$)^2$ difference 
between the lowest NSNS and RR modes, and thus does not perceive the 
additional winding energy. 

The appropriate field theory limits to D2- and D3-branes correspond to 
string regimes with and without tachyons respectively. Although the signs of 
the leading UV/IR mixing effects were not able to distinguish both regimes, 
their functional dependence on $\p$ was, which we could related to a closed 
string propagator. We found IR finite corrections for
D3-branes and IR divergent ones for D2-branes. 
Therefore the example of the D2-branes provides further support for
a correspondence between closed string tachyons and noncommutative 
instabilities associated to IR divergences. 
For D3-branes a weaker but still interesting result holds.
UV/IR mixing effects are still governed by the lowest modes in the associated
closed string sector. Moreover destabilizing UV/IR mixing effects are absent
for those sectors related to closed strings which are stable for any
radius of the twisted circle. 

For the noncommutative gauge theory on D3-branes both at orbifolds
and twisted circle backgrounds, the coefficients that governed the leading 
UV/IR effects could be directly derived from the matter content of the 
theory. It is worth stressing that this is not the case for the theory
on the transversal D2-branes, in which the spectrum is supersymmetric
while supersymmetry is broken by interactions. Yet the leading UV and IR 
behavior of the field theory can be understood in terms of closed 
strings. 

Since the severe consequences of UV/IR mixing were first recognized in 
\cite{Minwalla:1999px}, the different behavior of the planar and 
nonplanar sectors of the theory has been a puzzling fact. 
The planar sector can have UV divergences, while
the nonplanar translate UV into IR behavior. Cancellation of the
leading UV divergences for supersymmetric theories implies the
absence of the leading IR divergences. 
This suggests that planar and non-planar terms are not disconnected but 
they are part of a unified structure. In this line, we have shown that 
the gauge invariant effective action containing the leading UV/IR
mixing effects, reproduces also the leading contribution to
the planar graphs. 
The 1-loop vacuum graphs and planar 2-point functions can be 
obtained from the actions (\ref{saction}) and (\ref{strexch}) for the 
D3- and D2-brane theories respectively, when at least one of the 
open Wilson line operators is substituted by its component in the 
identity. 

For the gauge theory on the D3-branes, both planar and nonplanar pieces are 
finite. We can then smoothly send the noncommutativity parameter $\theta$
to zero. The action (\ref{saction}) does not trivialize in this limit. 
In the absence of noncommutativity it might look surprising that the 
closed exchange technique is able to reproduce some pieces of the 
effective action. This however points towards the fact that the  
field theory limit retains relevant information about closed strings. 
We regard this as a nontrivial implication of our results. 

In \cite{Armoni:2001uw} it was shown that the closed exchange technique
could also reproduce the logarithmic IR divergences of 4d noncommutative
gauge theories. However no string interpretation was given of the 
coefficients governing these divergences. It will be very interesting to 
determine whether a relation with closed string properties also holds for 
subleading UV/IR mixing effects.

\vspace{5mm}

{\centerline {\large \bf Acknowledgments}}

\vspace{3mm}

I would like to thank C. G\'omez, K. Landsteiner and 
specially A. Uranga for useful discussions.

\vspace{5mm}

\section{Appendix}

Let us consider a field with momentum $p_y=(m-{\bar \nu})/R$ 
($m \in \mathbf{Z}$) along a circle of radius $R$, with ${\bar \nu}$
arbitrary. We define $N_{\bar \nu}$ as
\beq
\sum_{m=-\infty}^\infty {1 \over (m-\bar \nu)^2 + R^2 p^2}=
{\pi \over R \, p} \, N_{\bar \nu}(R \, p)  \, .
\label{dis}
\eeq
For bosons with integer moding along the circle (${\bar \nu}=0$),
$N_0=1+2 n_B$, with $n_B$ the Bose-Einstein distribution; for 
fermions with half-integer moding (${\bar \nu}=1/2$), 
$N_{1 \over 2}=1-2 n_F$, with $n_F$
the Fermi-Dirac distribution. For arbitrary $\bar \nu$ we can
consider ${1 \over 2} (-1)^F (N_{\bar \nu}-1)$, with $F$ the fermion 
number, the 
generalization of $n_B$ and $n_F$. The sum (\ref{dis}) can be
evaluated in a closed form, with the result
\beq
N_{\bar \nu}(x) = {\sinh 2 \pi x \over \cosh 2 \pi x -
\cos 2 \pi {\bar \nu}}  \, .
\label{N}
\eeq 
In general ${1 \over 2}(-1)^F (N_{\bar \nu}-1)$, contrary to $n_{B,F}$,
is not always positive and thus it can not have the interpretation
of occupation numbers. This is not a problem, since $y$ will be 
always an spatial direction for us.

We focus next in a 4d euclidean noncommutative $U(1)$ theory with 
one direction compactified on a circle. The matter content 
will consist of two real scalars in the adjoint
representation: $\varphi$, with arbitrary ${\bar \nu}$, and $\phi$, 
with ${\bar \nu}=0$ and couplings as if obtained 
from the dimensional reduction of the gauge field from 5 to 4 dimensions. 
We want to derive the contribution to the polarization tensor of the gauge 
field and self-energy of $\phi$, from the 1-loop exchange of $\varphi$. 
We are only interested in the contributions that arise from high loop 
momenta. These are given by
\beqa
{\bar \Pi}^{AB} &=& 4g^2 {1 \over 2 \pi R} \sum_m \int {d^3l \over (2 \pi)^3} 
{2 L^A L^B-\delta^{A B}L^2 \over L^4} \sin^2  \label{pigen}
{\p .l \over 2} \, , \\ 
{\bar \Sigma} &=& - 4g^2 {1 \over 2 \pi R} \sum_m \int {d^3l \over (2 \pi)^3} 
{1 \over L^2} \sin^2 {\p .l \over 2} \, , 
\label{inte}
\eeqa
where the indices $A,B$ run over the non-compact directions
$\mu=1,2,3$ and the circle; $L=(l_\mu,l_y)$, with $l_y=(m-{\bar \nu})/R$, 
and the external momentum is $P=(p_\mu,p_y)$.
For $p_y \neq 0$, (\ref{pigen}) does not fulfills the Ward identities
and subleading contributions must be taken
into account. In the following we limit ourselves to $p_y=0$.

Using (\ref{dis}), we obtain
\beqa
{\bar \Pi}^{\mu \nu} &=& 2 g^2 \int {d^3 l \over (2 \pi)^3} \left[ \left(
{N_{\bar \nu} \over l}- {d N_{\bar \nu} \over dl} \right) 
{\hat l}^\mu {\hat l}^\nu -{N_{\bar \nu} \over l} \delta^{\mu \nu} \right] 
\sin^2 {\p.l \over 2} \, , \\
{\bar \Pi}^{yy} & = & 
2 g^2\int {d^3 l \over (2 \pi)^3}{d N_{\bar \nu} \over dl} 
\sin^2 {\p.l \over 2} \, , \\
{\bar \Sigma} & = &  - 2 g^2\int {d^3 l \over (2 \pi)^3}{N_{\bar \nu} \over l} 
\sin^2 {\p.l \over 2} \, ,
\eeqa
where ${\hat l}^\mu=l^\mu / l$. Using $\sin^2(x)={1 \over 2}(1-\cos 2x)$,
we can separate the planar and nonplanar pieces in the previous expressions.  
The integration over the angular variables
can be performed in an straightforward way, with the following results
for the non-planar part
\beqa
\Pi^{\mu \nu} &=& {\p^\mu \p^\nu \over \p^2} {g^2 \over 2 \pi^2 \p}
\int_0^\infty dl N_{\bar \nu} \, ( \sin \p l - \p l \cos \p l ) \, , 
\label{apinp1}\\
\Pi^{yy} &=& {g^2 \over 2 \pi^2 \p}
\int_0^\infty dl N_{\bar \nu} \, ( \sin \p l + \p l \cos \p l ) \, , \\
\Sigma &=& {g^2 \over 2 \pi^2 \p}
\int_0^\infty dl N_{\bar \nu} \, \sin \p l \, .
\label{apinp3}
\eeqa
The planar part is
\beq
{\hat \Pi}^{\mu \nu}=0 \, , \;\;\;\;\; 
{\hat \Pi}^{yy} =2 {\hat \Sigma} = 
-{g^2 \over \pi^2} \int_0^\infty dl N_{\bar \nu}\, l \, .
\label{apip}
\eeq

The expressions above apply when the field circulating in the loop
is a real scalar. For a general field with fractional momentum
${\bar \nu}$, we should multiply (\ref{apinp1})-(\ref{apip}) by
the number of degrees of freedom of the field times $(-1)^F$,
with $F$ its fermion number. When there are several adjoint scalars,
the self-energies are $\Sigma^{ab}=\delta^{ab} \Sigma$ with
$\Sigma$ as in (\ref{apinp1})-(\ref{apip}). When there are several
group factors, the non-planar 2-point functions will mix different
group labels if there are bifundamental fields running in the loop.
Each bifundamental degree of freedom gives half the contribution 
in (\ref{apinp1})-(\ref{apinp3}). Since (\ref{apinp1})-(\ref{apip})
only depend on the components of $p_\mu$ along the noncommutative
plane, the Wick rotation along the additional noncompact direction to 
Minkowski signature is trivial.

\end{document}